\documentclass[11pt]{article}
\usepackage{epsfig}
\usepackage{bm}
\usepackage{amssymb}
\usepackage{mathrsfs}
\usepackage{amsmath}
\usepackage{dsfont}

\begin{document}
\thispagestyle{empty}
\vspace*{0mm}

\begin{center}

{\LARGE 
Spectroscopy in finite density lattice field theory: \vskip3mm An exploratory study in the relativistic Bose gas} 
\vskip12mm
Christof Gattringer, Thomas Kloiber
\vskip5mm
Karl-Franzens University Graz \\
Institute for Physics\\ 
Universit\"atsplatz 5 \\
A-8010 Graz, Austria 
\vskip25mm

\end{center}

\begin{abstract}
We analyze 2-point functions in the relativistic Bose gas on the lattice, i.e., a charged scalar $\phi^4$ field with chemical potential $\mu$. Using a generalized worm algorithm we perform a Monte Carlo simulation in a 
dual representation in terms of fluxes where the complex action problem is overcome. We explore various aspects of lattice spectroscopy at finite density and zero temperature, such as the asymmetry of forward and backward propagation in time and the transition into the condensed phase. It is shown that after a suitable subtraction the exponents for forward and backward propagation are independent of $\mu$ and agree with the mass obtained from the propagator at $\mu = 0$. This holds for $\mu < \mu_c$ and shows that below the condensation transition the mass is independent of $\mu$ as expected from the Silver Blaze scenario.
\end{abstract}
\vskip10mm
\begin{center}
{\sl To appear in Physics Letters B.}
\end{center}

\vskip15mm
\noindent
{\tt christof.gattringer@uni-graz.at \\
thomas.kloiber@gmx.at }

\setcounter{page}0
\newpage
\noindent
\subsection*{Introduction}
In recent years several interesting developments for the simulation of lattice field theories at finite density were presented (for examples related to this work see, e.g., \cite{examples1,examples2,philipsen,wolff,korzec,aarts1,scorzato,phi4_1}). A key ingredient in these developments is a reformulation of the original lattice field 
theory in terms of new variables (dual variables) which usually are loops or surfaces. The dual variables are subject to constraints, but can be updated very efficiently with generalizations of the Prokof'ev-Svistunov worm algorithm \cite{worm}. Although some of the most interesting problems, such as finite density lattice QCD, are not yet under control, many interesting new concepts and techniques now enlarge our toolbox for such future challenges.

In this letter we address the problem of spectroscopy in finite density lattice field theory. We study 
2-point functions for a charged scalar $\phi^4$ field with a chemical potential (relativistic Bose gas). We use a dual representation \cite{phi4_1} in terms of fluxes where the partition function is a sum over closed loops of fluxes with real and positive weights, i.e., the complex action problem of the conventional representation is solved in the dual form. Our Monte Carlo update for the dual degrees of freedom is a generalization of the worm algorithm \cite{worm} and was presented in \cite{phi4_1}. 

Lattice spectroscopy is based on 2-point functions where suitable monomials of the lattice fields are separated in Euclidean time. At finite density these two-point functions are modified since the chemical potential introduces an asymmetry in time: Matter which is propagating forward in time is enhanced by the chemical potential $\mu$ over anti-matter propagating backward in time. Thus one expects that finite 
$\mu$ introduces an asymmetry in 2-point functions. Furthermore, in finite density field theory condensation transitions can appear when the chemical potential exceeds a critical value $\mu_c$ (see
\cite{aarts1,phi4_1} for an analysis of condensation in the relativistic Bose gas on the lattice). 
In this exploratory study we analyze the propagators in both phases, at $\mu < \mu_c$ where we study the asymmetry of the 2-point functions, and at $\mu > \mu_c$ where we observe condensation.

\subsection*{Conventional and dual formulation}

Since later we will discuss also a continuum analysis of the charged scalar field, we start with presenting the action in the continuum form,
\begin{equation}
S \; = \; \int d^{\, 4} x \, \Big[ \; |\partial_\nu \phi|^2 + (m^2 - \mu^2) |\phi|^2 + \lambda |\phi|^4
+ \mu(\phi^*\partial_4\phi - \partial_4 \phi^* \phi) \; \Big] \; .
\label{continuumact}
\end{equation}
The chemical potential $\mu$ couples to the conserved charge of the scalar field, but also shifts the mass parameter. This has an important implication for the free theory ($\lambda = 0$): for 
$\mu > m$ the action is unbounded from below and the theory becomes unstable. For finite $\lambda$, sufficiently large chemical potential flips the sign of the quadratic term, giving rise to the Mexican hat potential and the transition into the condensed phase.

In a lattice discretization the continuum action (\ref{continuumact}) assumes the form 
\begin{equation}
S \; = \; \sum_x \!\left( \eta |\phi_x|^2 + \lambda |\phi_x|^4 - 
\sum_{\nu = 1}^4 \left[ e^{-\mu \, \delta_{\nu,4} } \phi_x^\star \phi_{x+\widehat{\nu}}  \, + \,
e^{\mu \, \delta_{\nu,4} } \phi_x^\star \phi_{x-\widehat{\nu}}  \right]  \right) .
\end{equation}
The first sum is over the sites $x$ of a $N_s^3 \times N_t$ lattice with periodic boundary conditions, the second sum is over the four 
Euclidean directions $\nu = 1,2,3,4$, and $\widehat{\nu}$ denotes the unit vector in $\nu$-direction. The degrees of freedom are the 
complex valued field variables $\phi_x$ attached to the
sites $x$ of the lattice.  $\eta$ denotes the combination $8 + m^2$, where $m$ is the bare mass parameter and the $\phi^4$ coupling 
is denoted by $\lambda$. All parameters are in units of the lattice spacing $a$, in other words the 
lattice spacing is set to $a = 1$ throughout this paper. 
The partition sum $Z$ is obtained by integrating the Boltzmann factor $e^{-S}$ over all field configurations, 
$Z = \int D[\phi] e^{-S}$. The measure is a product over all lattice sites $x$, with $\phi_x$ being integrated over the complex plane,
i.e., $D[\phi] = \prod_x \int_\mathds{C} d \phi_x / 2\pi$.  

The lattice partition sum in the conventional form can be mapped exactly to a dual representation \cite{phi4_1}, where the new degrees of freedom are integer valued fluxes attached to the links. The dual partition sum is given by
\begin{eqnarray}
Z &\!\! = \!\!& \sum_{\{ k, l\}}  \left(\! \prod_x \delta\left( \sum_\nu \big[ k_{x,\nu}  -  k_{x-\widehat{\nu},\nu}  \big] \! \right) \right) \!\! \left( \prod_{x,\nu}\! \frac{1}{(|k_{x,\nu}| + l_{x,\nu})! \, l_{x,\nu}!} \right) 
\nonumber \\
& \!\! \times \!\!& \!\!
\left(\! \prod_x e^{\,- \mu k_{x,4} } \; 
W\!\left( \sum_\nu \big[ |k_{x,\nu}| +  |k_{x-\widehat{\nu},\nu}| + 2( l_{x,\nu} + l_{x-\widehat{\nu},\nu}) \big]  \right) \right)\! . 
\label{Zfinal}
\end{eqnarray}
The sum is over all configurations of two sets of flux variables:
$k_{x,\nu} \in \mathds{Z}$ and $l_{x,\nu} \in \mathds{N}_0$. 
The $k$-fluxes are subject to a constraint enforced by the product over Kronecker deltas (for notational convenience here 
denoted by $ \delta_{n,0} = \delta(n)$). The constraints enforce the conservation of $k$-flux at each site $x$, i.e., $\sum_\nu 
[ k_{x,\nu} - k_{x-\widehat{\nu},\nu}] = 0$ for all $x$. The $l$-fluxes are unconstrained.
The configurations of $k$- and $l$-fluxes are weighted with various weights, one of them including the chemical potential, 
and the factors $W$ are given by the integrals $W(n) = \int_0^\infty dr \, r^{n+1} \,  e^{-\eta r^2 - \lambda r^4}$, which can be 
evaluated numerically to arbitrary precision. It is obvious that also at finite $\mu$ the weight factors are real and positive, and 
the complex action problem is solved in the dual representation.

For the Monte Carlo update we use a generalized worm algorithm \cite{worm} that was presented in \cite{phi4_1}. In that
 paper thermodynamical observables such as the particle number $n$, the field expectation value 
$\langle |\phi|^2 \rangle$, and the corresponding susceptibilities were studied. All these observables can be obtained as derivatives of $\ln Z$ with respect to the parameters $\mu$ and $\eta$ and in the dual representation take the form of expectation values of moments of the flux variables. Based on these observables the phase diagram of the relativistic Bose gas and condensation phenomena were studied. 

\subsection*{2-point functions in the dual representation}

Using continuum notation the 2-point function needed for spectroscopy is given by 
\begin{equation}
C(t) \; = \; \int d^{\,3}y  \, \langle \phi(t,\vec{y}\,) \, \phi(0,\vec{0}\,)^* \rangle \; \propto \; e^{-E\,t} \; ,
\label{zeropprop}
\end{equation}
where we project to zero spatial momentum, such that the energy $E$ of the particle at rest, i.e., its mass, determines the exponential decay of $C(t)$. Thus, on the lattice we need to evaluate 2-point functions which in the conventional lattice representation are given by
\begin{equation}
\langle \phi_y \, \phi_z^* \rangle \; = \; \frac{1}{Z} \int D[\phi] \, e^{-S} \, \phi_y \, \phi_z^* 
\; \equiv \; \frac{1}{Z} \, Z_{y,z} \; .
\end{equation}
Here $Z_{y,z}$ denotes the lattice partition sum with two field insertions $\phi_y \, \phi_z^*$. It is straightforward to generalize the derivation in \cite{phi4_1} to find the dual representation of $Z_{y,z}$,
\begin{eqnarray}
Z_{y,z} \!&\!\!\! = \!\!\!&\! \sum_{\{ k, l\}} \! \left(\! \prod_x \delta\!\left(\!\sum_\nu \big[ k_{x,\nu}  -  k_{x-\widehat{\nu},\nu}  \big] - \delta_{x,y} + \delta_{x,z} \! \right)\!\!\right) \!\! \left( \prod_{x,\nu}\! \frac{1}{(|k_{x,\nu}| + l_{x,\nu})! \, l_{x,\nu}!}\!\right) 
\nonumber \\
\!\!\! \times & \!\!\!\!\!\! & \!\!\!\!\!\!\!\!\!\!
\left(\! \prod_x e^{\,- \mu k_{x,4} } \; 
W \! \left( \!\! \sum_\nu\!\big[ |k_{x,\nu}| \! + \!  |k_{x-\widehat{\nu},\nu}| + 2( l_{x,\nu} + l_{x-\widehat{\nu},\nu}) \big]  + \delta_{x,y} + \delta_{x,z} \right)\!\!\right)\! . 
\label{Zxy}
\end{eqnarray}
At the positions $y$ and $z$ where the source fields $\phi_y \, \phi_z^*$ are located the index of the weight factors $W(n)$ is increased (see the Kronecker deltas in the argument of the $W$). Also the constraints for the 
$k$-fluxes change to 
\begin{equation}
\prod_x \delta\!\left(\!\sum_\nu \big[ k_{x,\nu}  -  k_{x-\widehat{\nu},\nu}  \big] - 
\delta_{x,y} + \delta_{x,z} \! \right) \; .
\end{equation}
In the presence of the source fields $\phi_y \, \phi_z^*$ the usual local flux conservation 
$\sum_\nu \big[ k_{x,\nu}  -  k_{x-\widehat{\nu},\nu}  \big] = 0$ at all sites $x$ is modified such that the site $z$ constitutes a 
source of flux, the site $y$ a sink. Thus in the dual representation for $Z_{y,z}$ the set of configurations of allowed $k$-flux 
consists of closed loops and a single open line of flux connecting the sites $y$ and $z$.

To evaluate $Z_{y,z}$ efficiently we follow the strategy of \cite{korzec} and define a generalized partition function ${\cal Z}$  where we sum over all possible positions of the field insertions,
\begin{equation}
{\cal Z} \; = \; \sum_{u,v} Z_{u,v} \; .
\end{equation}
The configurations that constitute ${\cal Z}$ consist of closed loops of flux plus a single flux line with open ends at arbitrary sites $u,v$ of the lattice. For the Monte Carlo simulation of ${\cal Z}$ we can thus re-use the worm algorithm of \cite{phi4_1}, where now every step of the worm constitutes an admissible configuration of ${\cal Z}$ (not only the configurations where the worm has closed that contribute to $Z$). The two point functions are then obtained as
\begin{equation}
\langle \phi_y \, \phi_z^* \rangle \; = \; \frac{Z_{y,z}}{Z} \; = \;
\frac{ \langle \, \delta_{u,y} \, \delta_{v,z} \, \rangle_{\cal Z} }{ \langle \, \delta_{u,v} \;
W(f_u)/W(f_u + 2) \, \rangle_{\cal Z}} \; ,
\label{2pointflux}
\end{equation}
where $\langle .. \rangle_{\cal Z}$ denotes the expectation value with respect to the enlarged ensemble ${\cal Z}$. By $f_u = \sum_\nu\!\big[ |k_{u,\nu}| \! + \!  |k_{u-\widehat{\nu},\nu}| + 2( l_{u,\nu} + l_{u-\widehat{\nu},\nu}) \big] $ we denote the combined $k$- and $l$-flux at site $u$ that enters the weights $W$, and the 
reweighting with $W(f_u)/W(f_u + 2)$ in (\ref{2pointflux}) 
is necessary to correctly obtain the original partition sum $Z$ in the denominator of (\ref{2pointflux}). 
When taking into account the projection to zero momentum we end up with the following dual representation of the correlator $C(t)$:
\begin{equation}
C(t) \; = \; \frac{ \langle \, \delta_{t,u_4 - v_4} \, \rangle_{\cal Z} }{ \langle \, \delta_{u,v} \;
W(f_u)/W(f_u + 2) \, \rangle_{\cal Z}} \; .
\label{2pointfluxzero}
\end{equation}

In our simulations we use $N_s^3 \times N_t$ lattices with $N_s$ ranging from 16 to 32, and  $N_t$ from 32 to 100. A full 
update sweep is the combination of one local Metropolis sweep for the $l_{x,\nu}$ variables and one worm for the $k_{x,\nu}
$ fluxes. We typically use between $5\times10^5$ and $10^6$ measurements ($10^6$ in the plots we show) separated by 5 full update sweeps. The number 
of equilibration sweeps is typically $2.5\times10^4$. All errors we quote are statistical errors determined with the jackknife method. 
The weights $W$ in (\ref{Zfinal}) were evaluated numerically using Mathematica and pre-stored for the Monte Carlo 
simulation.

\subsection*{Analysis of the free case} 

We begin with an analysis of the free case, i.e., we set $\lambda = 0$. In this case we can compute the correlators $C(t)$ exactly using Fourier transformation. The purpose of studying the free case is twofold: The exact analytic result allows us to test the dual approach and the correct implementation of the two point function, but also serves to understand the general behavior of two point functions at non zero 
chemical potential. That latter aspect is most transparent in the continuum which we discuss explicitly (for the comparison to the results from the dual simulation we will of course use Fourier transformation on the lattice).

A straightforward calculation gives the continuum correlator $C(t)$ as the following Fourier integral over the 4-component of the momentum,
\begin{equation}
C(t) \; = \; \int_{-\infty}^\infty \frac{dp_4}{2\pi} \; \frac{e^{i p_4 t}}{p_4^2 + m^2 - \mu^2 + i2\mu \, p_4} \; .
\label{contprop}
\end{equation} 
Factorizing the denominator one can make explicit the poles in the complex $p_4$-plane: $p_4^2 + m^2 - \mu^2 + i2\mu \, p_4 = [p_4 - i(m-\mu)][p_4 + i(m+\mu)]$. Thus we have two poles on the imaginary axis at $i(m - \mu)$ and at $-i(m + \mu)$. For vanishing chemical potential the poles are equidistant from the origin at $\pm  i m$. Finite chemical potential $\mu$ shifts both poles downwards on the imaginary axis.
The integral (\ref{contprop}) can easily be computed using the residue theorem. For forward propagation, i.e., for $t > 0$ one closes the integration contour in the upper half of the complex plane and finds $C(t) =  \exp(-(m-\mu)t)/2m$, while for negative time $t < 0$ one finds  $C(t) = \exp((m + \mu)t)/2m$. For finite $T$ the temporal extent becomes finite with length $\beta = 1/T$. The corresponding propagator is obtained by making the $\beta = \infty$ propagator periodic,
\begin{equation}
C_\beta (t) \; \; = \; \; \sum_{n \in \mathds{Z}} C(t + n \beta) \; \; = \; \; \frac{1}{2m} \left[\frac{e^{-(m-\mu)t}}{1-e^{-(m-\mu)\beta}} \; + \;
 \frac{e^{-(m+\mu)(\beta - t)}}{1-e^{-(m+\mu)\beta}} \right] ,
\label{propperiod} 
\end{equation}
where the last expression was obtained by inserting the above results for forward and backward propagation into the sum and summing up the emerging geometric series. Thus for finite temperature the propagator is a superposition of a forward propagating part with an exponent of $m - \mu$ and a piece running backward in time with exponent $m + \mu$. 

Note that the result (\ref{propperiod}) is valid only for $\mu < m$, since for $\mu > m$ the action (\ref{continuumact}) at $\lambda = 0$  
is unbounded from below and the path integral does not exist. For the poles in the complex plane this instability corresponds to a transition of the pole $i(m - \mu)$ into the lower half of the complex plane.

\begin{figure}[t!]
\centering
\hspace*{-8mm}
\includegraphics[width=100mm,clip]{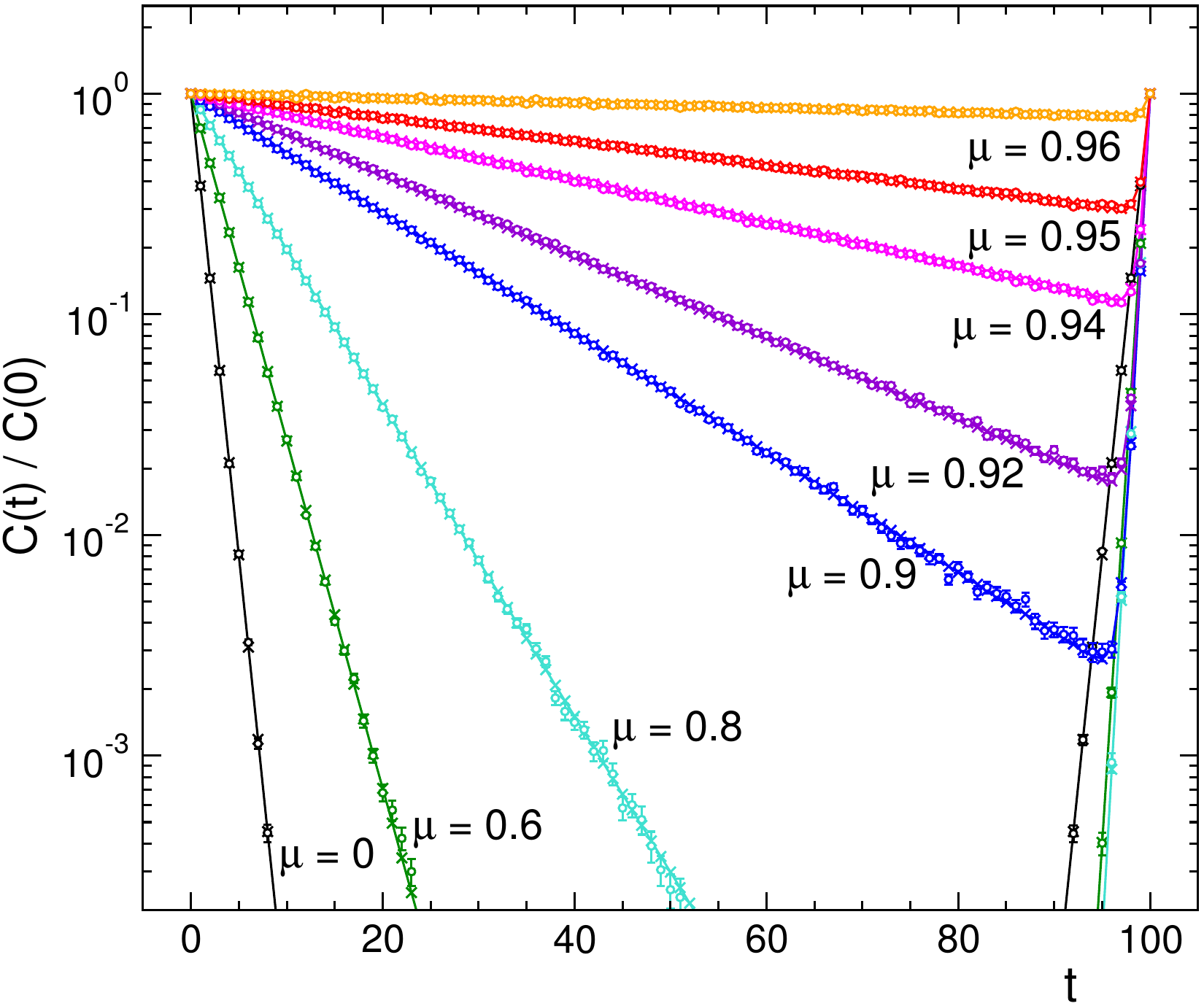}
\caption{The normalized propagator $C(t)/C(0)$ for the free case at $\eta = 9.0$ (i.e., $m = 1.0$) on $32^3 \times 100$ lattices. 
We show the results at  different  values of $\mu$ and compare the Monte Carlo data (circles)
to the propagator as computed from (lattice) Fourier transformation (crosses).}
\label{plot_free}
\end{figure}

Let us now come to the discussion of our numerical results for the free propagator and their comparison to lattice Fourier transformation. We are interested in studying the situation at low temperature and use $N_t = 100$ where the temperature in lattice units is $T = 1/N_t = 0.01$, much smaller than any other scale in the problem. The spatial volume is
$32^3$ and we work at $\lambda = 0$, $\eta = 9$, i.e., the mass is $m = 1.0$. This is the upper limit of the chemical potential $\mu$ we can consider before the theory becomes unstable.
In Fig.~\ref{plot_free} we show the propagator $C(t)/C(0)$ (i.e., normalized to 1 at $t = 0$) as a function of $t$ for different values of $\mu$. 
Circles are used for the results from the dual simulation, and crosses for the data from lattice Fourier transformation. 

The propagators show  the asymmetry between forward and backward propagation as expected from the continuum case we discussed. It is obvious that the dual Monte Carlo data and the Fourier results agree very well. This shows that the implementation of the propagator evaluation based on the generalized worm algorithm is correct and that we can reliably determine 2-point functions also at finite chemical potential.

\subsection*{Non zero quartic coupling} 

We now turn on the quartic coupling. In this case the path integral exists also for large values of the chemical potential and we can study the change of the propagators across the transition into the condensed phase. We  
work on lattices of size $32^3 \times 100$ and present results for $\lambda = 1.0$ and $\eta = 7.44$. For these parameters the condensation transition was found \cite{phi4_1} at a critical chemical potential of  $\mu_c = 0.170(1)$. For $\mu < \mu_c$ thermodynamical observables were found \cite{phi4_1} to be independent of $\mu$, a property sometimes referred to as Silver Blaze behavior \cite{silverblaze}. 

\begin{figure}[t!]
\centering
\hspace*{-8mm}
\includegraphics[width=100mm,clip]{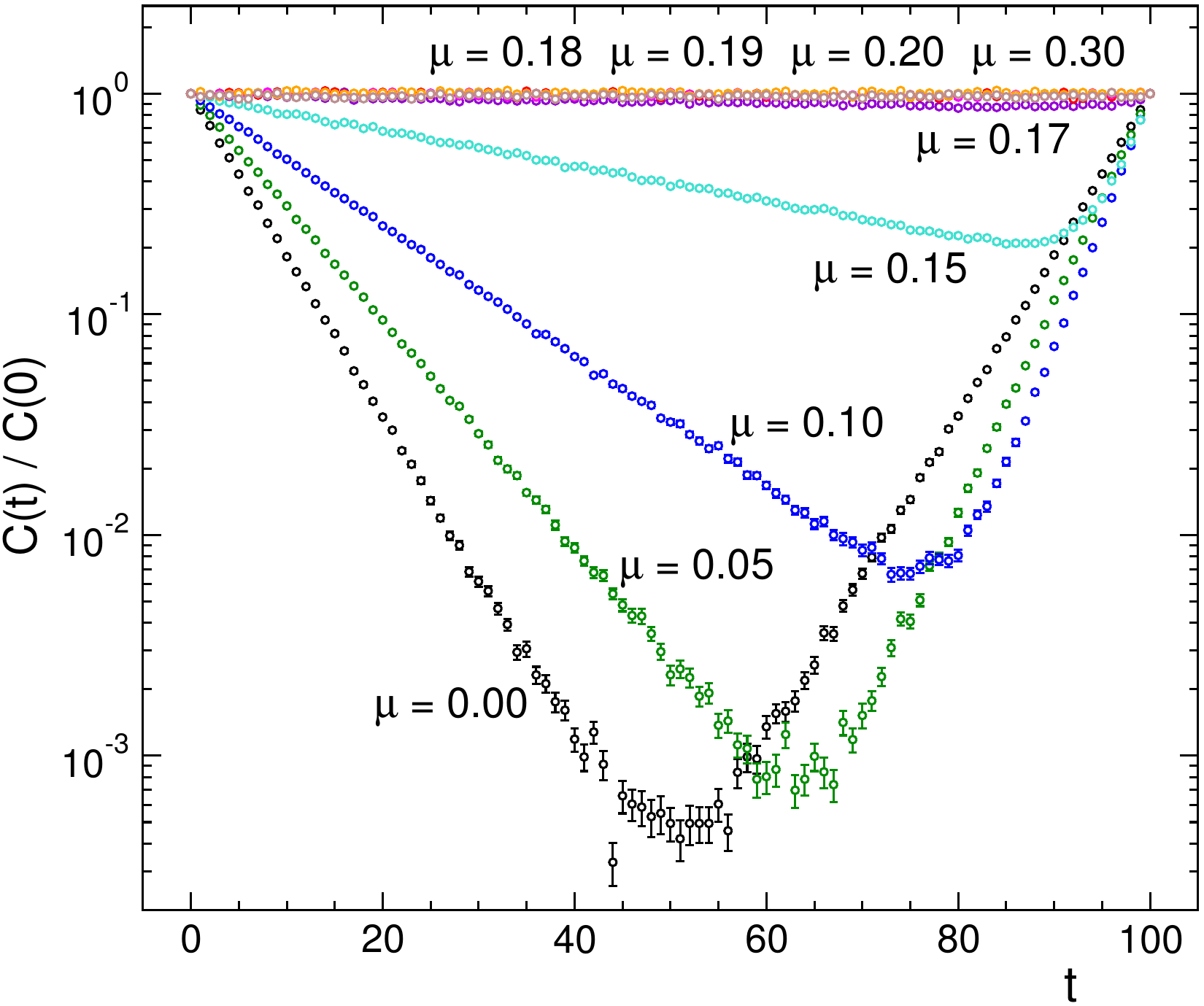}
\caption{The normalized propagator $C(t)/C(0)$ at $\eta = 7.44$ and $\lambda  = 1.0$ on lattices of size $32^3 \times 100$ for  different  values
of $\mu$.  }
\label{plot_full}
\end{figure}

\begin{table}[t]
\centering
\begin{tabular}{ccccc}
\hline
\(\mu \quad\)  &   \(E_{+}\)  &  \(E_{-}\)  &  \(E_{+}+\mu\)  &   \(E_{-}-\mu\) \\ 
\hline
0.00 \hspace{3mm} &  \hspace{3mm} 0.1687(3) \hspace{3mm} & \hspace{3mm} 0.1684(3)    \hspace{3mm} & \hspace{3mm} 0.1687(3) \hspace{3mm} & \hspace{3mm} 0.1684(3) \\
0.05 \hspace{3mm} &  \hspace{3mm} 0.1183(2) \hspace{3mm} & \hspace{3mm} 0.2178(3)    \hspace{3mm} & \hspace{3mm} 0.1683(2) \hspace{3mm} & \hspace{3mm} 0.1678(3) \\
0.10 \hspace{3mm} &  \hspace{3mm} 0.0685(2) \hspace{3mm} & \hspace{3mm} 0.267(1)    \hspace{3mm} & \hspace{3mm} 0.1685(2) \hspace{3mm} & \hspace{3mm}  0.167(1) \\
0.15 \hspace{3mm} &  \hspace{3mm} 0.0196(3) \hspace{3mm} & \hspace{4mm} -----  \hspace{2mm} & \hspace{3mm} 0.1696(3) \hspace{3mm} & \hspace{3mm}  ----- \\
\hline
\end{tabular}
\caption{Fit results for the exponents $E_+$ and $E_-$ for forward and backward propagation at different values of the chemical potential $\mu$  at $\eta = 7.44$ and $\lambda  = 1.0$ on lattices of size $32^3 \times 100$. 
In the last two columns we add (subtract) the chemical potential $\mu$ from the fit values $E_+$ and $E_-$.}
\label{fittable}
\end{table}

In Fig.~\ref{plot_full} we show the normalized propagator $C(t)/C(0)$ as a function of $t$ and compare the results for different values of the chemical potential $\mu$. Again, as expected, the propagator is symmetrical for $
\mu = 0$, while for $\mu > 0$ we observe a difference between forward and backward propagation. As in the free case we find very clean exponential decay showing up as straight lines due to the logarithmic scale used on 
the  vertical axis. For forward propagation the slope decreases with increasing $\mu$, while it increases for the backward running part.  The slope for forward propagation vanishes as the chemical potential reaches the 
critical value $\mu_c$ where the propagator becomes constant. For values of $\mu$ above $\mu_c$, i.e., in the condensed phase, the propagator remains constant.

Next we analyze if the slopes of forward and  backward propagation can be understood quantitatively. We determine the exponents $E_+$ and $E_-$ for forward and backward propagation from fits according to 
(\ref{zeropprop}) for several values of the chemical potential $\mu$. The results are listed in the second and third column of Table~\ref{fittable}. According to the Silver Blaze scenario \cite{silverblaze}
one expects that the mass of the 
excitation that $C(t)$ couples to is invariant for different values of $\mu < \mu_c$.  To check this expectation we add $\mu$ to $E_+$ and subtract it from $E_-$. The results for these reduced exponents 
at different $\mu < \mu_c$ should be constant in $\mu$ and agree with the mass obtained from the propagator at $\mu = 0$. The fourth and last columns in Table~\ref{fittable} show that this is indeed the case:  
The numbers for the reduced exponents are all close to 0.169, thus coinciding with the exponent of the propagator at $\mu = 0$
and agree well with the critical value $\mu_c = 0.170(1)$ determined in \cite{phi4_1}. We conclude that the propagators are understood also quantitatively and the mass of the lowest 
excitation indeed shows the Silver Blaze property, i.e., it is independent of $\mu$ up to $\mu_c$.

Let us finally comment in more detail on the behavior in the condensed phase: There the correlator should be dominated by the Goldstone mode which gives rise to algebraic decay, while the massive mode decays exponentially. 
However, at finite volume there is no spontaneous symmetry breaking and a finite volume Monte Carlo simulation only produces an ''erratic continuous drift through the degenerate vacuum states'' \cite{Hasenfratz1}. 
Two successful solutions to the problem are to either introduce a small external field to explicitly break the symmetry \cite{Hasenfratz2} or to use interpolators in the 2-point functions where $\phi_x$ is projected onto the 
dominant direction of the magnetization $M = \sum_x \phi_x$ for each configuration individually \cite{Hasenfratz1}. In both cases one needs the magnetization $M$. The problem with the dual representation we currently 
use is that we can only access observables that are even in the field $\phi$, while $M$ is odd. A more general dual representation of the model with magnetic field is possible where additional sets of dual variables, 
so called monomers, are introduced. The dual configurations are open strings with monomers at their endpoints. We hope that with this generalized dual representation we can successfully address the problem of studying the massless and massive modes of the condensed phase in the near future.

\subsection*{Summary and discussion}
In this work we have analyzed 2-point functions for a charged scalar $\phi^4$ field at finite density using the lattice formulation. The complex action problem of the standard form was 
overcome by mapping the partition sum to a flux representation where only real and non-negative terms appear. The 2-point functions were mapped to the dual formulation and we evaluated them 
at low temperatures and finite $\mu$ in a Monte Carlo simulation based on a generalized worm  algorithm.

In a first step we studied the free case to check the correctness and accuracy of the dual approach to 2-point functions. We observe the asymmetry between forward and backward propagation as expected from the 
analysis of the free case in the continuum and find perfect agreement of the dual simulation with the result from lattice Fourier transformation. 

For the case of non zero quartic coupling we again find the asymmetry of the propagators. As $\mu$ approaches its critical value $\mu$ the exponential decay of forward propagation weakens and vanishes at the 
condensation transition. In the condensed phase the propagator becomes constant. We find that after adding (subtracting) $\mu$ from the exponent $E_+$ ($E_-$) for forward (backward) propagation we get a number 
independent of $\mu$ which coincides with the mass obtained from the propagator at $\mu = 0$. This holds for all $\mu < \mu_c$ and shows that in this region the mass is independent of $\mu$ as expected from the Silver 
Blaze scenario. 

The analysis presented in this study of course only has an exploratory character, but we expect that the findings (and maybe some of the techniques) are relevant also for analyzing other systems with a Silver Blaze 
behavior, such as QCD at finite $\mu$ and low temperatures. There the chemical potential couples to quarks and the lowest excitation is the nucleon. Repeating the analysis of the free case for fermions, one again finds the 
difference between forward and backward propagation with exponents $m \pm \mu$, and it is expected that also in the coupled case after a suitable subtraction a mass independent of $\mu$ is obtained. While an analysis of this behavior is out of reach in full finite density lattice QCD, it would be interesting to check for this behavior of nucleon propagators 
in the QCD related model \cite{philipsen} obtained from hopping and strong coupling expansion.

\vskip10mm
\noindent
{\bf Acknowledgments:}
\vskip2mm
\noindent
We thank Gert Aarts, Shailesh Chandrasekharan, Ydalia Delgado Mercado and Hans Gert Evertz
for discussions. T.K. is supported by the FWF DK "Hadrons in Vacuum, Nuclei and Stars". This work was also supported by the DFG SFB TRR 55, "Hadron Physics from Lattice QCD".

\end{document}